\documentclass{llncs}

\usepackage{epsfig,amsmath,amsfonts}
\usepackage[dvips]{color}

\newcommand{\llaves}[1]{\left\{ #1 \right\}}
\newcommand{\la}{\langle}
\newcommand{\ra}{\rangle}
\newcommand{\assig}{\leftarrow}

\title{A distributed approximation algorithm for the minimum
  degree minimum weight spanning trees}

\author{Christian Lavault\inst{1} \and Mario
  Valencia-Pabon\inst{2}}

\institute{LIPN, CNRS UMR 7030, Universit\'e Paris-Nord \\ Av. J.-B.
  Cl\'ement, 93430 Villetaneuse, France \\email:
  lavault@lipn.univ-paris13.fr \and Departamento de Matem\'aticas,
  Universidad de los Andes \\ Cra. 1 No. 18A - 70, Bogot\'a, Colombia\\
  email: mvalenci@uniandes.edu.co}

\begin{document}

\maketitle

\begin{abstract}
Fischer \cite{FIS} has shown how to compute a minimum weight
spanning tree of degree at most $b \Delta^* + \lceil \log_b n\rceil$
in time $O(n^{4 + 1/\!\ln b})$ for any constant $b > 1$, where
$\Delta^*$ is the value of an optimal solution and $n$ is the number
of nodes in the network. In this paper, we propose a distributed
version of Fischer's algorithm that requires messages and time
complexity $O(n^{2 + 1/\!\ln b})$, and $O(n)$ space per node.\\

\end{abstract}
\section{Introduction}
Many computer communications networks require nodes to broadcast
information to other nodes for network control purposes; this is done
efficiently by sending messages over a spanning tree of the network.
Distributed minimum weight spanning tree algorithms are useful in
communication networks when one wishes to broadcast information from
one node to all other nodes and there with one cost assigned to each
channel in the network. If in such a minimum weight spanning tree the
degree of a node is large, it might cause an undesirable communication
load in that node. Therefore, the construction of minimum weight
spanning trees in which the degree of a node is the lowest possible is
needed. While it is easy enough to optimize the weight of a spanning
tree, it is often more difficult to satisfy constraints which also
involve the degrees of the nodes.  The problem of minimizing the
maximum degree of a spanning tree is known to be NP-hard, as the
Hamiltonian path problem is merely a special case of this problem
\cite{GJ}. In this paper, we consider the problem of finding a
distributed approximation algorithm for finding a minimum weight
spanning tree whose maximum degree is as low as possible.\\
{\it Previous and Related Work. }Let $\Delta^*$ be the degree of an
optimal solution. When edge weights are not considered, or assumed
uniform, a $\Delta^* + 1$ approximation algorithm for minimizing the
degree of spanning trees has been obtained by F\"urer and Raghavachari
\cite{FR}. A distributed version of the algorithm of F\"urer an
Raghavachari which maintains the same performance approximation
guarantee is proposed by Blin and Butelle \cite{BB}. For the weighted
case, Fischer \cite{FIS} gives an approximation algorithm that
computes a minimum weight spanning tree of degree at most $b \Delta^*
+ \lceil \log_b n \rceil$ in time $O(n^{4 + 1/\!\ln b})$ for any
constant $b >1$, which is the best-known algorithm for this problem up
to now. His algorithm is an adaptation of a local search algorithm of
F\"urer and Raghavachari \cite{FR} to the weighted case.  Recently,
Neumann and Laumanns
in \cite{NL} extend Fischer's algorithm to spanning forests.\\
{\it Our Results. }In this paper, we propose a distributed version of
Fischer's approximation algorithm that computes a minimum weight
spanning tree of degree at most $b \Delta^* + \lceil \log_b n \rceil$,
for any constant $b > 1$, where $n$ is the number of nodes of the
network and $\Delta^*$ is the maximum degree value of an optimal
solution.  Our distributed algorithm requires $O(n^{2 + 1/\!\ln b})$
messages and (virtual) time, and $O(n)$ space per node. From the
complexity analysis of our distributed algorithm, we are able to
derive that Fischer's sequential algorithm can be performed in $O(n^{3
  + 1/\!\ln b})$ time, which improves on Fischer's upper bound on the
time complexity.

The paper is organized as follows. In Section 2, we introduce the
model of computation and we present Fischer's sequential algorithm to
compute a minimum degree minimum weight spanning tree. In Section 3 we
describe our distributed algorithm and in Section 4 we prove its
correctness and complexity. Finally, Section 5 provides some
concluding remarks.
\section{Preliminaries}
{\it The model. } We consider the standard model of
asynchronous static distributed system. The point-to-point
communication network is associated a weighted undirected graph $G =
(V,E,w)$. The set of nodes $V$ represents the processors of the
network, the set of edges $E$ represents bidirectional non-interfering
communication channels operating between neighboring nodes, and $w$ is
a real-valued function defined on $E$, which represents a cost
assigned to each channel of the network. No common memory is shared by
the nodes (processes).  In such networks, any process can generate one
single message at a time and can send it to all its neighbors in one
time step. Every process owns one distinct identity from $\{1,\ldots,n\}$.
However, no node has a global knowledge of the network topology,
except of its own incident edges (e.g., every node is unaware of its
neighbors' identities).  The distributed algorithm is event-driven and
does not use time-outs, i.e. nodes can not access a global clock in
order to decide what to do. Moreover, each node runs the algorithm,
determining its response according to every type of message received.
Namely, the algorithm specifies for any one node which computation is
to be performed and/or which message be sent. The algorithm is started
independently by all nodes, perhaps at different times. When the
algorithm starts, each node is unaware of the global network topology
but of its own edges. Upon termination, every node knows its
neighbors' identities within an approximated minimum degree minimum
weight spanning tree. The efficiency of a distributed algorithm is 
evaluated in terms of {\it message}, {\it time} and {\it space} complexity 
as follows (see \cite{TEL}). The {\it message complexity} of a distributed 
algorithm is the total number of messages sent over the edges. 
We also assume that each message contains $O(\log n + R)$ bits, where 
$|V| = n$ and $R$ is the number of bits required to represent any real 
edge weight.
In practical applications, messages of such a kind are considered of
``constant" size. The {\it time complexity} is the total (normalized)
time elapsed from a change. The {\it space complexity} is the space
usage per node.

{\it The Problem. }Let $G = (V,E,w)$ be a real-weighted
graph modeling the communication network. A spanning tree $T =
(V_T,E_T)$ of $G$ is a tree such that $V_T = V$ and $E_T \subseteq E$.
The weight of a spanning tree $T$ of $G$ equals the sum of the weights
of the $|V| - 1$ edges contained in $T$, and $T$ is called a {\it
  minimum weight spanning tree}, or MWST, if no tree has a smaller
(minimum) weight that $T$.  Our goal is to find a distributed
polynomial algorithm to compute a MWST such that its maximum degree is
as low as possible.  We denote by $N_G(u)$ and $N_T(u)$ the set of
neighbors of node $u$ in $G$ and in $T$, respectively.

{\it Fischer's sequential algorithm. }
Let $T$ be a tree on $n$ nodes. Define the {\bf rank} of $T$ to be the
ordered $n$-tuple $(t_n,\ldots,t_1)$ where $t_i$ denotes the number of
nodes of degree $i$ in $T$. Define a lexicographical order on these
ranks; a tree $T'$ on $n$ nodes is of lower rank that $T$ if $t'_j <
t_j$ for some $j$ and $t'_i = t_i$ for $i = j+1,\ldots,n$.  Clearly,
when an edge is added to a spanning tree, it creates a cycle.
Conversely, removing any edge from the induced cycle results again in a
spanning tree.  A {\it swap} is defined to be any such exchange of
edges; a swap is said {\it cost-neutral} if the edges exchanged are of
equal weight. Consider a swap between the edges $xw \in T$ and $uv \not \in T$,
with $x \not\in \{u,v\}$. Such a swap may increase by one the degree
of both $u$ and $v$ in $T$, but it also reduces the degree of $x$.
So, the rank of $T$ is decreasing if the degree of $x$ in $T$
is at least the maximal degree of $u$ and $v$ plus 2.
A {\bf locally optimal} minimum weight spanning tree is a MWST in
which no cost-neutral swap can decrease the rank of the tree.

\begin{theorem}(Fischer \cite{FIS})
\label{tf1}
If $T$ is a locally optimal MWST, and $\Delta_T$ is the maximum degree
in $T$, then $\Delta_T \leq b\Delta^* + \lceil \log_bn \rceil$ for
any constant $b > 1$, where $\Delta^*$ is the maximum degree of an
optimal solution.
\end{theorem}
As can be deduced from the proof of Theorem \ref{tf1} in Fischer's
paper \cite{FIS}, in order to construct a locally optimal spanning tree,
it is sufficient to consider those nodes with degree at least equal
to $\Delta_T - \lceil \log_bn \rceil$ among high-degree nodes.
Fischer's polynomial algorithm to compute a locally optimal
spanning tree can be described as follows. \\

\noindent {\bf Fischer's algorithm :}
\begin{enumerate}
\item[0.] Start with any MWST $T$. Let $b>1$ be the desired
  approximation parameter. Let $l < n$ be the number of distinct edge
  weights, $w_1,\ldots,w_l$, in $T$.
\item[1.] Let $\Delta_T$ be the current maximum degree in $T$.
\item[2.] For every node $v \in G$, check for appropriate
  improvements. Conduct a depth first traversal of $T$ starting from
  $v$.
  \begin{enumerate}
  \item[2.1.] Let $w$ be the current vertex on the traversal of $T$,
  and $P$ be the $vw$-path in $T$. 
  \item[2.2.] Assign variables $M_1,\ldots,M_l$ such that $M_i$
  denotes the maximum degree of those nodes adjacent to edges of
  weight $w_i$ in $P$. 
  \item[2.3.] If there is an edge $vw \in G$, let $w_i$ be its
  weight. If $M_i$ is at least two greater that the degree of $v$ and
  $w$ in $T$, and $M_i \geq \Delta_T - \lceil \log_bn \rceil$, the the
  edge $uv$ can be used to reduce the high-degree rank of $T$. Conduct
  the appropriate swap on $T$, and repeat to Step (1) for the next
  round. 
  \item[2.4.] If no appropriate cost-neutral swap was found in any of
  the traversals, terminate.
  \end{enumerate} 
\end{enumerate}
Fischer proves in \cite{FIS} that each round of his previous algorithm
takes $O(n^2)$ time and that the number of rounds can be bounded by
$O(n^{2+1/\!\ln b})$. Therefore, Fischer's algorithm computes a
locally optimal minimum weight spanning tree in time $O(n^{4+1/\!\ln b})$.
\section{Description of the algorithm}
In Section~\ref{highlevel}, a high-level description of our
distributed algorithm is given, and in Section~\ref{detail}, we
detail the description of the algorithm. Every process is running
the following procedure, which consists of a list of the responses
to each type of messages generated. Each node is assumed
to queue the incoming messages and to reply them in First-Come,
First-Served order (FIFO). Any reply sent is completed before
the next is started and all incoming messages are also delivered
to each node via an initially empty First-Come First-Served queue.
\subsection{High-Level Description} \label{highlevel}
\begin{figure}[htp]
\centering
\epsfsize=0.5\hsize
\noindent
\epsfbox{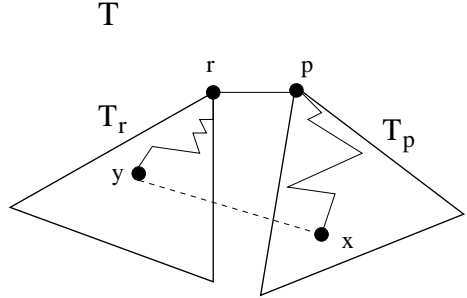}
\caption{Example of a cost-neutral swap between edges $pr \in T$ and
  $xy \not\in T$.}
\label{fig1}
\end{figure}

Let $G$ be a connected graph modeling an interconnection network.  We
now describe the general method used to construct a locally optimal
minimum weight spanning tree $T$ of $G$. First, we assume that some
current minimum weight spanning tree $T$ of $G$ is already
constructed.  (Various MWST distributed algorithms are $\Theta(|E| + n
\log n)$ message optimal, e.g.~\cite{AWE,GHS}, while the best time
complexity achieved is $O(n)$ in~\cite{AWE}.) Next, for each edge $pr
\in T$, let $T_r$ (resp. $T_p$) be the subtree of $T \setminus pr$
containing the node $r$ (resp. $p$) (see Fig.~\ref{fig1}).
\noindent The algorithm is divided into rounds, and each
round is consisting of the following four phases. \\

\noindent $\bullet$ {\bf Initialization phase. }Let $\Delta_T$ be the
maximum current degree of the MWST $T$. First, each node of $T$ must
know $\Delta_T$. This can be done by starting an election on $T$,
where all nodes broadcast their degree in $T$. Before checking an edge
to find an appropriate cost-neutral swap (see Fig.~\ref{fig1}), we
need to do the following. For each edge $pr$ in $T$ and for each node
$x \in T$, we determine, which of the nodes $p$ or $r$ is the closer
to $x$ (in terms of distance in $T$). For the purpose, the node $p$
initiates a broadcast by forwarding the token $\la pr \ra$ to
all nodes in $T_p$. When a node $x \in T_p$ receives the token
$\la pr\ra$, $x$ knows that for any appropriate swap involving
the edge $pr$, it needs to find a neighbor $y$ in $G$ such that $y \in
T_r$.
In an analogous way, node $r$ initiates a similar broadcast in $T_r$. 

\noindent $\bullet$ {\bf Search-edge phase. }Whenever an edge $pr$ in the
current MWST $T$ ends the Initialization phase, it begins searching
for an edge $xy \not\in T$ to perform an appropriate swap. Thus, node
$p$ starts a search for a node $x \in T_p$ with a neighbor $y \in
T_r$, for which $w(pr) = w(xy)$ and such that the maximal value of the
degrees of $p$ and $r$ in $T$ is at least two greater than the degree
of $x$ and $y$. If such a node $x$ is found, it sends a message to $p$
(via edges in $T_p$) to announce success. If there is no node in $T_p$
meeting the previous conditions then $p$ is informed that edge $pr$
can not be used to an appropriate swap on $T$. Similarly, node $r$
starts such a search in $T_r$. If no appropriate cost-neutral swap is
found in this phase, the algorithm terminates. 

\noindent $\bullet$ {\bf Edge-election phase. }If a node $p$ enters
this phase then, there exists a pair of edges $pr \in T$ and $xy
\not\in T$ which can execute an appropriate swap on $T$.  However, $p$
must ensure that it belongs to the one only pair of edges that can
achieve such a swap. So, $p$ starts an election procedure and it
chooses among the possible initiators (i.e. the nodes that reached
this phase) one node to perform the swap. So, each initiator $p$
forwards its identity to all other nodes in $T$. The elected node is
the one with the minimal identity.

\noindent $\bullet$ {\bf Edge-exchange phase. }If node $p$ wins the
Edge-election phase, then there exists again at least one pair of
edges $pr \in T$ and $xy \not\in T$ that can be used to complete the
appropriate swap on $T$ (reducing the high-degree rank of $T$).  Thus,
$p$ informs $r$ that they are not connected in $T$ anymore and starts
a search in $T_p$ for the node $x$ (assuming $x \in T_p$).  When $x$
is found, it informs $y$ that they are now connected within $T$.
Finally, $x$ sends a message to all other nodes in the current tree to
inform that a new round can initiate.
\subsection{Detailed description} \label{detail}
The algorithm is described for a node $p$ in the network.
We begin by describing the variables maintained by $p$. \\

\noindent {\bf Local variables of $p$: } (Let $T$ denote
the current minimum weight spanning tree.)

\begin{itemize}
\item {\it $\mbox{Neigh}_p[r]:\ \llaves{\mbox{branch, unbranch}}$.}
  Node $p$ maintains the variable for each $r \in N_G(p)$.
  An edge $pr$ belongs to $T$ if and only if
  $\mbox{\it Neigh}_p[r] = \mbox{\it branch}$.
\item {\it $d^T_p, \Delta_T, d_p[r]: \mbox{integer};\ w_p[r] : \mbox{real}$.}
  The variables $d^T_p$ and $\Delta_T$ denote the degree of $p$ and
  the maximal degree of $T$, respectively. For each $r \in N_G(p)$, the
  variables $d_p[r]$ and $w_p[r]$ denote the degree of the neighbor
  $r$ of $p$ in $T$ and the weight of the edge $pr$ in $G$, respectively.
\item {\it $\mbox{Father}_p[r] : \mbox{integer}$.} For each edge $ry
  \in T$, the variable $\mbox{\it Father}_p[r]$ denotes the neighbor
  of $p$ closer to node $r$ in $T_r$ (if $p \in T_r$). {\it Default
    value:} $\mbox{\it Father}_p[r] = p$, for each $r \in T$.
\item {\it $\mbox{Side}_p[uv] : \mbox{integer}$.} For each edge
    $uv \in T$, the variable $\mbox{\it Side}_p[uv]$ denotes the node
    $r \in \{u,v\}$ closer to $p$ in $T$. {\it Default value:}
    $\mbox{\it Side}_p[uv] = \mbox{\it udef}$, for each edge $uv \in T$.
\item {\it $\mbox{CountSide}_p[uv], \mbox{CountFail}_p[uv] : \mbox{integer}$.}
  For each edge $uv \in T$, the variables $\mbox{\it CountSide}_p[uv]$
  and $\mbox{\it CountFail}_p[uv]$ maintain the number of neighbors of $p$
  which already instanced the variable $\mbox{\it Side}_\_[uv]$.
  Moreover, the variable $\mbox{\it CountFail}_p[uv]$ counts the number
  of neighbors of $p$ which find no edge to replace $uv$. {\it Default values:}
  $\mbox{\it CountSide}_p[uv] = \mbox{\it CountFail}_p[uv] = 0$,
  for each edge $uv \in T$.
\item {\it $\mbox{EndInit}_p[r] : \llaves{0,1,2}$.} For each $r \in N_T(p)$,
  the variable maintains the state of the Initialization phase of
  the edge $pr \in T$. {\it Default value:}
  $\mbox{\it EndInit}_p[r] = 0$, for each $r \in N_T(p)$.
\item {\it $\mbox{EdgeFind}_p[uv] : \llaves{0,1}$.} For each edge $uv \in T$,
  the boolean variable is used to check whether an edge $xy \not\in T$
  is found in $G$ which replaces $uv$. {\it Default value:}
  $\mbox{\it EdgeFind}_p[uv] = 0$, for each edge $uv \in T$.
\item {\it $\mbox{NeighFail}_p, \mbox{\#Fails}_p : \mbox{integer}$.}
  The variable $\mbox{\it NeighFail}_p$ maintains the number of
  neighbors of $p$ in $T$ s.t. the associated edge in $T$ is useless
  to any edge-exchange in a current round. If
  $\mbox{\it NeighFail}_p = |N_T(p)|$, it means that no edge
  incident to $p$ can be used in view of an exchange in a current round.
  The variable $\mbox{\it \#Fails}_p$ maintains the number of nodes
  in $T$ for which there is no edge incident to the ones that can
  be used in an exchange. If in a round, $\mbox{\it \#Fails}_p = |T|$,
  it means that the algorithm is terminated. {\it Default values:}
  $\mbox{\it NeighFail}_p = \mbox{\it \#Fails}_p = 0$.
\item {\it $\mbox{Mode}_p : \llaves{\mbox{election, non-election}}$.}
  If during a Search-edge phase a subset of edges in $T$ finds an
  edge appropriate to an exchange, all corresponding edges-incident
  nodes begin an election to decide on the edge to be exchanged.
  If $p$ knows that an Election phase is running,
  {\it $\mbox{Mode}_p = \mbox{election}$}. {\it Default value:}
  $\mbox{\it Mode}_p = \mbox{\it non-election}$.
\item {\it $\mbox{State}_p : \llaves{\mbox{winner, loser, udef}}$.}
  The variable maintains the state of $p$ during an Edge-election phase
  ({\it winner or loser}). If {\it $\mbox{State}_p = \mbox{udef}$},
  it means that $p$ does not know if an Edge-election phase
  is taking place in a current round. {\it Default value:}
  $\mbox{\it State}_p = \mbox{\it udef}$.
\item {\it $\mbox{CountLoser}_p : \mbox{integer}$.}
  During an Edge-election phase, the variable maintains the number
  of nodes that lost the election w.r.t. node $p$. Clearly, if
  $\mbox{\it CountLoser}_p = |T| - 1$, node $p$ wins the election,
  i.e. an edge incident to $p$ becomes elected.
  {\it Default value:} $\mbox{\it CountLoser}_p = 0$.
\item {\it $\mbox{NodeElec}_p : \mbox{integer}$.}
  The variable maintains the identity of the winning node
  during an Edge-election phase.
  {\it Default value:} $\mbox{\it NodeElec}_p = \mbox{\it udef}$.
\item {\it $\mbox{EdgeElec}_p :
\mbox{integer}\times\mbox{integer}\times\mbox{integer}\times\mbox{integer}\times\mbox{integer}$.}
  The variable maintains the pair of elected edges (if any)
  in view of an appropriate exchange in a current round.
  If $\mbox{\it EdgeElec}_p = (d,p,r,x,y)$, then a pair of edges
  $pr$ and $xy$ is found in $G$, s.t. $pr \in T$, $xy \not\in T$,
  $\mbox{\it Side}_x[pr] = p$, $\mbox{\it Side}_y[pr] = r$, $w_x[y] =
  w_p[r]$, and where $d = \max\{d^T_p, d_p[r]\}$. {\it Default value:}
  $\mbox{\it EdgeElec}_p = \mbox{\it udef}$.
\end{itemize}
  
\noindent Now, each node $p$ executes the following steps.\\

\noindent {\bf Initial assumptions: }Assume that the algorithm starts 
with any MWST $T$ of $G$ already given. We need the algorithm
constructing $T$ to be ``process terminating'' (i.e. every node knows
that the MWST algorithm is terminated; however, no distributed
termination detection is available). So, we can assume that $p$ knows
its degree $d^T_p$ in $T$, and for each $r \in N_G(p)$, the variables
$Neigh_p[r]$, $d_p[r]$ and $w_p[r]$ are correctly computed. Let $b >
1$ be the desired approximation parameter. \\

\noindent {\bf Initialization phase :}
\begin{enumerate}
\item[1.] In order to determine the maximum degree $\Delta_T$ in the
  current tree $T$, $p$ initiates an election (all nodes in $T$ are
  initiators) by sending the token $\la d^T_p, p \ra$ to all
  its neighbors in $T$. Node $p$ wins the election if the token
  $\la d^T_p, p \ra$ is maximal w.r.t. the lexicographic
  order. As all vertices are initiators, we also use this step to
  initialize the remaining local variables of $p$ with their default
  values (see the definition of local variables of $p$ above).
\item[2.] For each edge $uv \in T$ and for each node $r \in T$,
  we need to compute the values of variables $\mbox{\it Side}_p[uv]$
  and $\mbox{\it Father}_p[r]$. This is done as follows.

\begin{enumerate}
\item[2.1.] \textbf{For} each $z \in N_T(p)$ \textbf{do}
      \begin{enumerate}
      \item[$\bullet$] $\mbox{\it Side}_p[pz] \textbf{$\assig$} p$;
      \item[$\bullet$] send $\la \mbox{\bf side},p,z,p\ra$ to each node
      $r \in N_T(p)$ s.t. $r \neq z$;
      \end{enumerate}

\item[2.2.] Upon receipt of $\la \mbox{\bf side},u,v,q\ra$
      \begin{enumerate}
      \item[$\bullet$] $\mbox{\it Side}_p[uv] \textbf{$\assig$} u$;
      \item[$\bullet$] $\mbox{\it Father}_p[u] \textbf{$\assig$} q$;
      \item[$\bullet$] \textbf{If} $p$ is not a leaf \textbf{then} send $\la \mbox{\bf side},u,v,p\ra$
      to each node $r \in N_T(p)$ s.t. $r \neq q$;
      \item[$\bullet$] \textbf{If} $p$ is a leaf \textbf{then} send $\la \mbox{\bf end-side},u,v\ra$
      to node $\mbox{\it Father}_p[u]$;
      \end{enumerate}

\item[2.3.] Upon receipt of $\la \mbox{\bf end-side},u,v\ra$
      \begin{enumerate}
      \item[$\bullet$] $\mbox{\it CountSide}_p[uv] \textbf{$\assig$} \mbox{\it CountSide}_p[uv] + 1$;
      \item[$\bullet$] \textbf{If} $p \neq u$ \textbf{and}
      $\mbox{\it CountSide}_p[uv] = |N_T(p)| - 1$ \textbf{then}
      send $\la \mbox{\bf  end-side},u,v\ra$ to node $\mbox{\it Father}_p[u]$;
      \item[$\bullet$] \textbf{If} $p = u$ \textbf{and}
      $\mbox{\it CountSide}_p[uv] = |N_T(p)| - 1$ \textbf{then} \\
      -\ send $\la \mbox{\bf check-side},u\ra$ to node $v$;

      -\ $\mbox{\it EndInit}_p[v] \textbf{$\assig$} \mbox{\it EndInit}_p[v] + 1$;

      -\ \textbf{If} $\mbox{\it EndInit}_p[v] = 2$ \textbf{then} edge $pv \in T$
      finishes the Initialization phase and can \textbf{go to} Step 3.
      \end{enumerate}

\item[2.4.] Upon receipt of $\la \mbox{\bf check-side},q\ra$
      \begin{enumerate}
      \item[$\bullet$] $\mbox{\it EndInit}_p[q] \textbf{$\assig$} \mbox{\it EndInit}_p[q] + 1$;
      \item[$\bullet$] \textbf{If} $\mbox{\it EndInit}_p[q] = 2$ \textbf{then}
      edge $pq \in T$ finishes the Initialization phase and can \textbf{go to} Step 3.
      \end{enumerate}
\end{enumerate}
\end{enumerate}

\noindent {\bf Search-edge phase:}
\begin{enumerate}
\item[3.] If there is some neighbor $q$ of $p$ in $T$ for which
  $\mbox{\it EndInit}_p[q] = 2$, it means that the edge $pq \in T$
  is ready to initiate a search for some possible unused edge $xy$
  of $G$ in view of an appropriate exchange. For our purpose,
  the edge $pq \in T$ must meet the condition $\max\{d^T_p, d_p[q]\} \geq %
  \Delta_T - \lceil \log_bn\rceil$. Otherwise, $pq$ can not be considered
  in this phase. This is done as follows.

  \begin{enumerate}
  \item[3.1.] \textbf{If} $\exists q \in N_T(p)$ s.t.
  $\mbox{\it  EndInit}_p[q] = 2$ \textbf{and} $\max\{d^T_p, d_p[q]\} \geq %
  \Delta_T - \lceil \log_bn\rceil$ \textbf{then} send
  $\la \mbox{\bf change},p,q,d_{pq},w_p[q]\ra$ to each node
  $r \in N_T(p)$, with $r \neq q$, and where $d_{pq} = \max\{d^T_p, d_p[q]\}$.

  \item[3.2.] Upon receipt of $\la \mbox{\bf change},u,v,d_{uv},w_u[v]\ra$

    \begin{enumerate}
      \item[$\bullet$] \textbf{If} $\exists r \in N_G(p)$ s.t.
      $\mbox{\it Neigh}_p[r] = \mbox{\it unbranch}, \mbox{\it Side}_r[uv] = v, %
      w_p[r] = w_u[v], \mbox{ and } d_{uv} \geq \max\{d^T_p, d_p[r]\} + 2$
      \textbf{then} send $\la \mbox{\bf find},u,v,p,r\ra$ to node
      $\mbox{\it Father}_p[u]$; \\
      \textbf{else} \\
      -\ \textbf{If} $p$ is a leaf \textbf{then} send
      $\la \mbox{\bf fail},u,v\ra$ to node $\mbox{\it Father}_p[u]$;\\
      -\ \textbf{If} $p$ is not a leaf \textbf{then} send
      $\la \mbox{\bf change},u,v,d_{uv},w_u[v]\ra$ to each node
      $r \in N_T(p)$ s.t. $r \neq \mbox{\it Father}_p[u]$;
    \end{enumerate}

  \item[3.3.] Upon receipt of $\la \mbox{\bf find},u,v,q,r\ra$

        \begin{enumerate}
        \item[$\bullet$]$\mbox{\it EdgeFind}_p[uv] \textbf{$\assig$} 1$;
        \item[$\bullet$] \textbf{If} $p \neq u$ \textbf{then} send
        $\la \mbox{\bf find},u,v,q,r\ra$ to node $\mbox{\it Father}_p[u]$;
        \item[$\bullet$] \textbf{If} $p = u$ \textbf{then} $p$ is ready
        to begin the Edge-election phase and can \textbf{go to} Step 4.
        \end{enumerate}

  \item[3.4.] Upon receipt of $\la \mbox{\bf fail},u,v\ra$

     \begin{enumerate}
     \item[$\bullet$] $\mbox{\it CountFail}_p[uv] \textbf{$\assig$} \mbox{\it CountFail}_p[uv] +
     1$;

     \item[$\bullet$] \textbf{If} $\mbox{\it EdgeFind}_p[uv] \neq 1$ \textbf{then} \\
     -\ \textbf{If} $p \neq u$ \textbf{and} $\mbox{\it CountFail}_p[uv] = |N_T(p)| - 1$
     \textbf{then} send $\la \mbox{\bf fail},u,v\ra$ to node $\mbox{\it Father}_p[u]$; \\
     -\ \textbf{If} $p = u$ \textbf{and} $\mbox{\it CountFail}_p[uv] = |N_T(p)| - 1$
     \textbf{then}
        \begin{enumerate}
        \item[ ] $\mbox{\it NeighFail}_p \textbf{$\assig$} \mbox{\it NeighFail}_p
        + 1$; 
        \item[ ] send $\la \mbox{\bf fail1},u,v\ra$ to node $v$;
        \item[ ] \textbf{If} $\mbox{\it NeighFail}_p = |N_T(p)|$ \textbf{then}
        send $\la \mbox{\bf end},p\ra$ to each node $r \in N_T(p)$ in order to
        check whether the algorithm is terminated;
        \end{enumerate}

     \end{enumerate}

  \item[3.5.] Upon receipt of $\la \mbox{\bf fail1},q,p\ra$
    or $\la \mbox{\bf end},r\ra$

    \begin{enumerate}
    \item[$\bullet$] $\mbox{\it \#Fails}_p \textbf{$\assig$} \mbox{\it \#Fails}_p
    + 1$;  
    \item[$\bullet$]\textbf{If} $\mbox{\it \#Fails}_p = |T|$
    \textbf{then} the algorithm terminates.
    \end{enumerate}

  \end{enumerate}

\end{enumerate}

\noindent {\bf Edge-election phase:}

\begin{enumerate}
\item[4.] This phase begins when $p$ receives a message $\la \mbox{\bf
    find},p,r,x,y\ra$, which means that the edge $pr \in T$ can be
  exchanged for an unused edge $xy \in G$. Before the change, $p$ must
  ensure that it is the only node in $T$ that performs this operation.
  To this end, $p$ initiates an election procedure in $T$ and it picks
  among the possible initiators (i.e. the nodes that reached this
  phase) the one that makes the edge exchange. Note that from this
  step till the end of the current round, any message of type $\la
  \mbox{\bf side}\ra$, $\la \mbox{\bf end-side}\ra$, $\la \mbox{\bf
    check-side}\ra$, $\la \mbox{\bf change}\ra$, $\la \mbox{\bf
    find}\ra$, $\la \mbox{\bf fail}\ra$, $\la \mbox{\bf fail1}\ra$,
  and $\la \mbox{\bf end}\ra$ received by $p$ is ignored. Let $d_{pr}
  = \max\{d^T_p, d_p[r]\}$.
    \begin{enumerate}

    \item[4.1.] \textbf{If} $\mbox{\it Mode}_p = \mbox{\it non-election}$ \textbf{then}

       \begin{enumerate}
       \item[$\bullet$] $\mbox{\it EdgeElec}_p \textbf{$\assig$} (d_{pr},p,r,x,y)$;
       $\mbox{\it NodeElec}_p \textbf{$\assig$} p$; $\mbox{\it Mode}_p \textbf{$\assig$} \mbox{\it election}$;

       \item[$\bullet$] Send $\la \mbox{\bf elec},p,p\ra$ to each node $q \in N_T(p)$;
       \end{enumerate}
       \textbf{else} ($\mbox{\it Mode}_p = \mbox{\it election}$)

       \begin{enumerate}
       \item[$\bullet$] \textbf{If} $\mbox{\it NodeElec}_p = p$
         \textbf{and}
         $\mbox{\it State}_p = \mbox{\it udef}$ \textbf{then} \\
         \ \ let $\mbox{\it EdgeElec}_p = (d,p,z,a,b)$; \textbf{If}
         $d_{pr} > d$ \textbf{then} $\mbox{\it EdgeElec}_p
         \textbf{$\assig$} (d_{pr},p,r,x,y)$;
       \end{enumerate}

    \item[4.2.] Upon receipt of $\la \mbox{\bf elec},q,z\ra$

      \begin{enumerate}
      \item[$\bullet$] \textbf{If} $\mbox{\it Mode}_p = \mbox{\it non-election}$
      \textbf{then} \\
      - $\mbox{\it NodeElec}_p \textbf{$\assig$} q$; $\mbox{\it Mode}_p \textbf{$\assig$} \mbox{\it election}$;
      $\mbox{\it State}_p \textbf{$\assig$} \mbox{\it loser}$; \\
      - Send $\la \mbox{\bf lost},p,p\ra$ to each node $r \in N_T(p)$; \\
      - \textbf{If} $p$ is not a leaf \textbf{then} send $\la \mbox{\bf elec},q,p\ra$
      to each node $r \in N_T(p)$ s.t $r \neq z$; \\
      \textbf{else} ($\mbox{\it Mode}_p = \mbox{\it election}$) \\
      - \textbf{If} $q < \mbox{\it NodeElec}_p$ \textbf{then}

         \begin{enumerate}
         \item[-] \textbf{If} $\mbox{\it State}_p \neq \mbox{\it loser}$
         and $\mbox{\it NodeElec}_p = p$ then \\
         \ $\mbox{\it State}_p \textbf{$\assig$} \mbox{\it loser}$; \\
         \ send $\la \mbox{\bf lost},p,p\ra$ to each node $r \in N_T(p)$;

         \item[-] \textbf{If} $p$ is not a leaf \textbf{then} send
         $\la \mbox{\bf elec},q,p\ra$ to each node $r \in N_T(p)$ s.t. $r \neq z$;

         \item[-] $\mbox{\it NodeElec}_p \textbf{$\assig$} q$;
         \end{enumerate}

      \end{enumerate}

    \item[4.3.] Upon receipt of $\la \mbox{\bf lost},q,t\ra$

      \begin{enumerate}
      \item[$\bullet$] \textbf{If} $\mbox{\it Mode}_p = \mbox{\it election}$
      \textbf{and} $\mbox{\it State}_p = \mbox{\it udef}$ \textbf{then} \\
      - \textbf{If} $p \neq q$ \textbf{then}
      $\mbox{\it CountLoser}_p \textbf{$\assig$} \mbox{\it CountLoser}_p + 1$. \\
      Moreover, if $\mbox{\it CountLoser}_p = |T| - 1$ \textbf{then}
      $\mbox{\it State}_p \textbf{$\assig$} \mbox{\it winner}$; \textbf{go to} Step 5. \\
      - \textbf{If} $p = q$ \textbf{then} $\mbox{\it State}_p \textbf{$\assig$} \mbox{\it loser}$;

      \item[$\bullet$] \textbf{If} $\mbox{\it State}_p \neq \mbox{\it winner}$
      \textbf{and} $p$ is not a leaf \textbf{then} send $\la \mbox{\bf lost},q,p\ra$
      to each node $r \in N_T(p)$ s.t. $r \neq t$.
      \end{enumerate}

    \end{enumerate}
\end{enumerate}

\noindent {\bf Edge-exchange phase:}
\begin{enumerate}
\item[5.] If $p$ wins the Edge-election phase, i.e. $\mbox{\it State}_p = \mbox{\it winner}$,
  with $\mbox{\it EdgeElec}_p = (d_{pr},p,r,x,y)$. Then the edge $pr \in T$
  is ready to be exchanged for the unused edge $xy \in G$.
  Note that from this step till the end of the current round, any message
  received by $p$ which is distinct from types $\la \mbox{\bf disconnect}\ra$,
  $\la \mbox{\bf connect}\ra$, $\la \mbox{\bf new}\ra$, and $\la \mbox{\bf nround}\ra$
  is ignored. So, during this phase the following is done.

  \begin{enumerate}
  \item[5.1.] $\bullet$ Send $\la \mbox{\bf disconnect},r,p\ra$ to node $r$; \\
              $\bullet$ $\mbox{\it Neigh}_p[r] \textbf{$\assig$} \mbox{\it unbranch}$;
              $d_p[r] \textbf{$\assig$} d_p[r] - 1$; $d^T_p \textbf{$\assig$} d^T_p -1$; \\
              $\bullet$ Send $\la \mbox{\bf connect},p,r,x,y\ra$
              to each node $t \in N_T(p) : t \neq r$;

  \item[5.2.] Upon receipt of $\la \mbox{\bf disconnect},p,q\ra$ \\
    $\bullet$ $\mbox{\it Neigh}_p[q] \textbf{$\assig$} \mbox{\it unbranch}$;
    $d_p[q] \textbf{$\assig$} d_p[q] - 1$; $d^T_p \textbf{$\assig$} d^T_p -1$;

  \item[5.3.] Upon receipt of $\la \mbox{\bf connect},u,v,x,y\ra$

     \begin{enumerate}
     \item[$\bullet$] \textbf{If} $p \neq x$ and $p$ is not a leaf
     \textbf{then} send $\la \mbox{\bf connect},u,v,x,y\ra$ 
     to each node $t \in N_T(p) : t \neq \mbox{\it Father}_p[u]$;

     \item[$\bullet$] \textbf{If} $p = x$ \textbf{then} \\
     -\ send $\la \mbox{\bf new},y,p\ra$ to node $y$ (via the 
     unused edge $py \in G$); \\
     -\ $\mbox{\it Neigh}_p[y] \textbf{$\assig$} \mbox{\it branch}$;
     $d_p[y] \textbf{$\assig$} d_p[y] + 1$; $d^T_p \textbf{$\assig$} d^T_p + 1$;
     \end{enumerate}

  \item[5.4.] Upon receipt of $\la \mbox{\bf new},p,q\ra$ \\
    $\bullet$ $\mbox{\it Neigh}_p[q] \textbf{$\assig$} \mbox{\it branch}$;
    $d_p[q] \textbf{$\assig$} d_p[q] + 1$; $d^T_p \textbf{$\assig$} d^T_p + 1$; \\
    $\bullet$ send $\la \mbox{\bf nround},p\ra$ to each node
    $t \in N_T(p)$;

  \item[5.5.] Upon receipt of $\la \mbox{\bf nround},z\ra$ \\
    $\bullet$ \textbf{If} $p$ is not a leaf \textbf{then} send
    $\la \mbox{\bf nround},p\ra$ to each node $w \in N_T(p)$ s.t. 
    $w \neq z$; \\
    $\bullet$ \textbf{Go to} step 1. (This round is ended and $p$ 
    executes a new round.)
  \end{enumerate}
\end{enumerate}
\section{Correctness and Complexity}
\begin{theorem} \label{tcor}
The distributed algorithm described in Section 3 for computing a locally
optimal minimum weight spanning tree is correct and deadlock-free.
\end{theorem}

\begin{proof}
Let $T$ be the MWST computed in a current round of the algorithm.
If any node $p$ of $T$ meets the condition $\mbox{\#Fails}_p = |T|$
(the number of nodes in $T$), it means that no edge in $T$ can find
another edge outside $T$ in view of a cost-neutral swap to reduce
the rank of $T$. In such a situation, the algorithm terminates.
By definition, $T$ is indeed a locally optimal minimum weight
spanning tree and the algorithm is correct. Moreover, by construction,
when a node of $T$ generates a message to send through the edges in $T$,
it uses each edge of $T$ at most once; and it is removed (i.e. disappears)
either by reaching its destination or whenever it arrives to a leaf
of $T$. Upon receipt of a message, any node $p$ executes some
instructions if $p$ and the information within the message meet
specific properties. Otherwise, the message is ignored by the node.
Finally, when a cost-neutral swap reducing the rank of $T$ is found,
all nodes in $T$ are informed and initiate a new round (see Steps
5.4 and 5.5 ). Otherwise, all nodes in $T$ are informed that the
algorithm is terminated (see Steps 3.5 and 3.6). Consequently,
we may conclude that the algorithm is also deadlock-free. \qed
\end{proof}

\begin{lemma} \label{lcom}
Each round of the distributed algorithm in Section 3 requires
$O(n)$ messages and time $O(n)$.
\end{lemma}

\begin{proof}
  We assume that the algorithm starts with any MWST $T$ of $G$
  available beforehand ($|T| = |V| = n$).  In order to compute the
  maximum degree of the current tree $T$, an election is triggered in
  Step~(1). The time and message complexity of such a tree election is
  $O(n)$ (see \cite{TEL}).  Now, consider any edge $uv \in T$. By
  construction, both nodes $u$ and $v$ generate one single message of
  each type $\la \mbox{\bf side}\ra$, $\la \mbox{\bf change}\ra$ and
  $\la \mbox{\bf connect}\ra$ (see Steps 2.1, 3.1 and 5.1, resp.),
  which traverses only once each edge in $T_u$ and $T_v$,
  respectively.  These messages disappear whenever they reach a leaf
  node. Therefore, the number of such messages sent over $T$ is at
  most equal to $3(n-2)$. Next, consider the edges of $T$ that are
  incident to nodes in $T_u$.  One of the nodes incident to such edges
  generates one message of each type $\la \mbox{\bf end-side}\ra$,
  $\la \mbox{\bf find}\ra$ and $\la \mbox{\bf fail}\ra$ (see Steps 2.2
  and 3.2, resp.), which is sent to node $u$. Thus, these messages
  traverse each edge in $T_u$ only once and the same property holds in
  the subtree $T_v$. Therefore, the number of such messages sent over
  $T$ is at most equal to $3(n-2)$. Further, each node in $T$ can
  generate one message of each type $\la \mbox{\bf elec}\ra$, $\la
  \mbox{\bf lost}\ra$ and $\la \mbox{\bf nround}\ra$ (see Steps 4.1,
  4.2 and 5.4, resp.), which is broadcast through the tree $T$.
  Therefore, the number of such messages sent over $T$ is at most
  equal to $3(n-1)$. Finally, The remaining messages are exchanged
  once between neighbors.  Therefore, the number of messages sent over
  $T$ within a current round is $O(n)$. The time complexity of the
  algorithm during a round can be derived from the above analysis by
  observing that a node $u$ can broadcast a message to all its
  neighbors in one time step and the distance in $T$ from $u$ to any
  destination node is at most $n-1$. \qed
\end{proof}

\begin{lemma} \label{lround}
  For any constant $b>1$, the number of rounds used by the distributed
  algorithm in Section 3 can be bounded from above by $O(n^{1 +
    1/\!\ln b})$.
\end{lemma}

\begin{proof}
  We use a potential function similar to Fischer's, with the
  difference that only high-degree nodes have a potential value
  greater or equal to one. A similar potential function is also used
  in~\cite{NL}. Let $\Delta_T$ be the maximum degree of the current
  MWST $T$ on $n$ nodes during a round of the algorithm.  Let $\delta
  = \max\{\Delta_T - \lceil \log_b n\rceil, 0\}$ and let $d^T_v$ be
  the degree of a node $v$ in $T$. The {\it potential} of the node $v$
  is define as follows,
\begin{displaymath}
\phi_v = \left\{\begin{array}{ll}
          e^{d^T_v - \delta} & \quad \textrm{if }\ \ d^T_v \geq
          \Delta_T - \lceil \log_b n\rceil, \\
          1/2 & \quad \textrm{otherwise.} \\
          \end{array}\right.
\end{displaymath}
Denote $\Phi_T = \sum_{v \in T}\phi_v$. Since $b > 1$, 
$\Phi_T \leq n e^{\lceil \log_b n\rceil} \leq n e^2 n^{1/\!\ln b}$.

\noindent Let us now compute the change in potential, $\Delta\Phi$,
when the algorithm performs a local improvement involving a node of
degree at least $\Delta_T - \lceil \log_b n\rceil$ in $T$.  Assume
edge $xy$ is added to $T$ and edge $uv$ is removed from $T$; also
assume w.l.o.g. that $d^T_u \geq d^T_v$. Since the algorithm only
performs swaps which reduce the degree of some high-degree node, we
have that $d^T_u \geq \Delta_T - \lceil \log_b n\rceil$ and $d^T_u
\geq \max\{d^T_x, d^T_y\} + 2$. By a simple case analysis, it is easy
to check that, in a local improvement, the potential decreases of the
smallest amount possible if $d^T_u = \Delta_T - \lceil \log_b n\rceil$
and $d^T_v, d^T_x, d^T_y < \Delta_T - \lceil \log_b n\rceil$.  In such
a case, any local improvement reduces the potential by $1/2$.
Therefore, in any local improvement, $\Delta\Phi \geq 1/2$.  This
implies that after at most two local improvements (i.e. rounds),
$\Phi_T$ decreases of at least one unit. Hence, the algorithm finds a
locally optimal MWST in $O(n^{1 + 1/\ln b})$ rounds. \qed
\end{proof}

\begin{theorem}
  For any constant $b > 1$, the distributed algorithm in Section 3
  requires $O(n^{2 + 1/\!\ln b})$ message and time, and $O(n)$ space
  per node to compute a minimum weight spanning tree of maximum degree
  at most $b \Delta^* + \lceil \log_b n\rceil$, where $n$ is the
  number of nodes of the network and $\Delta^*$ is the maximum degree
  value of an optimal solution.
\end{theorem}

\begin{proof}
  The algorithm is assumed to start with any MWST $T$ of $G$ (e.g. by
  using the algorithm in \cite{AWE} beforehand).  Now, for each edge
  $uv \in T$, each node $p$ maintains the variables $\mbox{\it
    Side}_p[uv]$, $\mbox{\it CountSide}_p[uv]$, $\mbox{\it
    CountFail}_p[uv]$ and $\mbox{\it EdgeFind}_p[uv]$ in particular
  (see Section~\ref{detail}). Since $T$ is a tree on $n$ nodes, its
  number of edges is $n-1$, and so the algorithm requires $O(n)$ space
  per node. Finally, by Theorems~\ref{tf1} and \ref{tcor}, and
  Lemmas~\ref{lcom} and \ref{lround}, the theorem follows. \qed
\end{proof}
Note that the proof of Lemma \ref{lround} works also in the sequential
case. Therefore, we obtain the following corollary, which improves on 
Fischer's time complexity.
\begin{corollary}
  For any constant $b > 1$, Fischer's sequential algorithm in~\cite{FIS} 
  (Section 2) finds a minimum weight spanning tree of maximum degree 
  at most $b \Delta^* + \lceil \log_b n\rceil$ in $ O(n^{3 + 1/\!\ln
  b})$ time.
\end{corollary}
\section{Concluding Remarks}
In the paper, we present a distributed approximation algorithm 
which computes a minimum weight spanning tree of degree at most 
$b \Delta^* + \lceil\log_b n \rceil$, for any constant $b > 1$. 
The message and time complexity of the algorithm is 
$O(n^{2 + 1/\!\ln b})$ and it requires $O(n)$ space per node. 
To our knowledge, this is the first distributed approximation 
algorithm for the minimum degree minimum weight spanning tree problem.


%
\section*{Appendix}
In order to facilitate the reading of our paper, we give the proof of
Fischer's Theorem \ref{tf1} in Section 2 as
follows.\\

\noindent {\bf Proof of Theorem \ref{tf1} (Fischer~\cite{FIS}).}
Let $b > 1$ be any constant and let $G$ be a connected
graph on $n$ nodes. Consider a locally optimal MWST T of $G$ with
maximum degree $\Delta_T$. Let $S_i$ denote the set of nodes of degree
at least $i$ in $T$. Clearly, $|S_{\Delta_T}| \geq 1$. Since $|S_i|
\leq n$ for all $i$, the ratio $|S_{i-1}|/|S_i|$ can not be greater
that $b$ for $\log_b n$ consecutive values of $i$. Therefore, for any
constant $b >1$, there exists some integer $\delta$ in the range
$\Delta_T - \lceil \log_b n \rceil \leq \delta \leq \Delta_T$ such
that $|S_{\delta - 1}|/|S_{\delta}| \leq b$. Suppose we choose an
integer $\delta$ to satisfy this property, and remove from $T$ the
edges adjacent to nodes in $S_{\delta}$. Let $T_{\delta}$ denote the
remaining edges of $T$. As $T$ is initially connected, then there are
at least $\delta|S_{\delta}| +1 - (|S_{\delta}| -1)$ or $(\delta
-1)|S_{\delta}| +2$ connected components in $T_{\delta}$.

Consider the graph $G_{\delta}$ formed by contracting every component
of $T_{\delta}$. Since any MWST of $G$ must contain a MWST of
$G_{\delta}$, any MWST must include at least $(\delta
-1)|S_{\delta}| +1$ edges from $G_{\delta}$.

Consider an edge $vw \in G$ not in $T$ between two components of
$T_{\delta}$. Let $P^T$ denote the $vw$-path in $T$, and
$P^T_{\delta}$ denote those edges of $P^T$ which appear in
$G_{\delta}$, the edges on the path which are adjacent to any node in
$S_{\delta}$. Suppose neither $v$ nor $w$ is in $S_{\delta - 1}$.
Since $T$ is locally optimal, no cost-neutral swap can reduce the rank
of $T$, so $vw$ must be more expensive that any edge in
$P^T_{\delta}$. Since $vw$ and $P^T_{\delta}$ form a cycle in
$G_{\delta}$, this implies that $vw$ may not participate in a MWST of
$G_{\delta}$. Therefore, only edges which are adjacent to nodes in
$S_{\delta - 1}$ may participate in a MWST of $G_{\delta}$, and any
MWST of $G$ must contain at least $(\delta -1)|S_{\delta}| +1$ edges
that are adjacent to $S_{\delta - 1}$.

Earlier we chose $\delta$ to satisfy the inequality $|S_{\delta -
  1}|/b \leq |S_{\delta}|$. Substituting, we see there must be at
least $((\delta -1)|S_{\delta - 1}|/b) + 1$ edges adjacent to nodes in
$S_{\delta-1}$. Therefore, the average degree of a node in
$S_{\delta-1}$ must be at least $\frac{(\delta -1)|S_{\delta -1}|
  +b}{b|S_{\delta -1}|}$, and so, $\Delta^* > \frac{\delta - 1}{b}$.

Combining this with the possible range for $\delta$, we find that
$\Delta_T \leq b\Delta^* + \lceil \log_b n\rceil$.  \qed
\end{document}